\documentclass[]{article}

\usepackage{amsmath,amssymb,float,geometry,pslatex} 
\usepackage{graphicx,natbib}
\usepackage{authblk}

\def\bes{\begin{eqnarray}}\def\ees{\end{eqnarray}}\def\beq{\begin{equation}}\def\eeq{\end{equation}}

\def\nn{\nonumber} \def\d{\textrm d} \def\ba#1\ea{\begin{align}#1\end{align}}
\def\bsa#1#2\esa{\begin{subequations}\label{#1} \begin{align}#2\end{align} \end{subequations}}

\def\lp{\left(} 
\def\rp{\right)}

\def\k{{\bf{k}}}

\allowdisplaybreaks

\def\m{\omega}

\def\e{\textrm{e}}

\def\D{\mathcal{D}}

\def\bes{\begin{eqnarray}}\def\ees{\end{eqnarray}}\def\beq{\begin{equation}}\def\eeq{\end{equation}}

\def\nn{\nonumber}
\def\d{\textrm d}
\def\ba#1\ea{\begin{align}#1\end{align}}

\def\bsa#1#2\esa{\begin{subequations}\label{#1}
\begin{align}#2\end{align} \end{subequations}}

\def\g{\nabla}\def\f{\frac}\def\b{\mathbf}

\begin{document}

\title{Dominant Resonance in Parametric Subharmonic Instability of Internal Waves}

\author[1]{Y. Liang}
\author[2]{L.-A. Couston}
\author[3]{Q. C. Guo}
\author[1,3]{M.-R. Alam\thanks{reza.alam@berkeley.edu}}

\affil[1]{Applied Science and Technology, University of California Berkeley, Berkeley, CA 94720}
\affil[2]{CNRS, Aix Marseille Univ, Centrale Marseille, IRPHE, Marseille, France}
\affil[3]{Department of Mechanical Engineering, University of California Berkeley, Berkeley, CA 94720}
\date{}
\maketitle

\begin{abstract}

Parametric Subharmonic Instability (PSI) is one of the most important mechanisms that transfer energy from tidally-generated long internal waves to short steep waves. Breaking of these short waves results in diapycnal mixing through which oceanic abyssal stratification is maintained.  
It has long been believed that PSI is strongest between a primary internal wave and perturbative waves of half the frequency of the primary wave. Here, we rigorously show that this is not the case. Specifically, we show that neither the initial growth rate nor the maximum long-term amplification occur at the half frequency, and demonstrate that the dominant subharmonic waves have much longer wavelengths than previously thought.

\end{abstract}

\section{Introduction}

Internal gravity waves are ubiquitous in world's density-stratified oceans. They mainly arise from barotropic tides flowing over topographic features, or wind disturbing the upper ocean's mixed layer \cite[][]{wunsch2004}. Internal waves transport energy over long distances in oceans and when break result in considerable mixing which contributes to, e.g., oceanic circulations through lifting cold water from the ocean basin \citep{Garrett2003a}, and the lives of a wide range of ocean creatures by redistributing nutrients \citep{Boyd2007,Harris1986}.

The underlying mechanism(s) that lead to the breaking of internal waves is not yet fully understood despite significant recent progresses made in our understanding of such waves. There is a nearly general consensus that low mode internal waves, such as those generated by tides, need to somehow transfer their energy to shorter waves which are steeper and hence more prone to breaking. Several mechanisms for such transfer of energy from long to short waves have been put forward, among them interaction with topographic features \citep{Lamb2014,Sarkar2017,Swinney2014} and parametric subharmonic instability \citep[e.g.][]{Hasselmann1967,Davis1967,Dauxois2013} are the most important ones.

Parametric Subharmonic Instability (PSI) is the instability of a primary internal wave to two lower-frequency internal waves with (initially) infinitesimal amplitudes thus \textit{disturbances}. This happens if the primary and the two disturbance waves' frequencies and wavenumber vectors (respectively $\omega_{0,1,2}$ and $\k_{0,1,2}$) satisfy the triad resonance condition \citep[e.g.][]{Lombard1996,Karimi2014,Koudella2006}
\ba\label{007}
\omega_0=\omega_1+\omega_2,~~\k_0=\k_1+\k_2.
\ea
Through this instability, energy is transferred from the primary wave to the two disturbance waves. The flow of energy may continue until the amplitudes of the disturbance waves, initially infinitesimal, become of the same order as the amplitude of the primary wave or even higher, and that is why the disturbance waves are said to be \textit{resonated}. This resonance is not through an explicit external force but instead is caused by what appears to be \emph{parameters} in the governing equations, hence it is called \textit{parametric} resonance or instability\footnote{Some literature \citep[e.g.][]{Dauxois2018}, to be precise, call this resonance a ``Parametric Subharmonic instability or PSI'' only if the disturbance waves have a frequency half of the frequency of primary wave. If the disturbance waves have a different frequency, then it is called ``Triadic Resonant Instability or TRI''.}. Resonated subharmonic waves obtained through PSI usually have smaller vertical and horizontal scales than the primary wave. Thus, PSI constitutes a pathway for energy transfer to steeper waves which are more prone to breaking. 

PSI was first studied \citep[in 1960s, e.g.][]{Hasselmann1967,Davis1967} as a subset of the general wave resonance theory \citep{Phillip1981,Marcus2009}, and has since been reported in several field studies \citep[e.g.][]{Alford2008b,Chinn2012a}. The current understanding of PSI is based on a linear stability theory \citep[established in the 1970s, e.g.][]{Martin1972,McEwan1971}. Assuming that the amplitude of the primary wave is constant, and that amplitudes of disturbance waves are much smaller than the primary wave, the linear stability theory predicts an exponential growth rate for a pair of perturbing waves that satisfy the resonance conditions \eqref{007} \citep[see e.g.][]{Martin1972,Koudella2006,Bourget2013}.

An internal wave, however, can undergo parametric subharmonic instability simultaneously with a countably infinite pairs of subharmonic waves. In order to accurately determine the role of PSI on the evolution of oceanic internal wave spectrum as well as to answer how efficiently an internal wave can transfer its energy to smaller scales, it is critical to know which specific pair (or pairs) of disturbance waves draw the most energy from the primary internal wave. In other words, it is critical to know which triads are resonated the strongest out of all the different triad resonance possibilities. The classical linear stability theory assumes that these infinite resonance possibilities are independent (or decoupled) and predicts that the pair of subharmonic waves with half of the frequency of the primary wave has the largest growth rates and thus is expected to dominate in the process of PSI \citep[e.g.][]{Staquet2002,Bourget2013}.

However, laboratory and numerical experiments on PSI do not support the predominance of half frequency resonant waves. For example, in a series of experiments on PSI in a wave tank $21$m long, $1.2$m deep and filled with linearly stratified fluid, Martin et al. \citep{Martin1972} obtained multiple subharmonic waves generated from PSI of a mode-three internal wave, but none of them were at the half frequency. In another attempt, Joubaud et al. \citep{Joubaud2012} sends a horizontally propagating mode-one internal wave at frequency $0.95N$ ($N$: Brunt-V{\"a}is{\"a}l{\"a} frequency) but observes two subharmonic waves not at the half frequency but at frequencies $0.38N$ and $0.57N$. Also, direct simulation of internal beams generated by a tidal flow (frequency $\omega_0$) over bottom topography results in strongest subharmonic waves at frequencies $0.4\omega_0$ and $0.6\omega_0$ \citep{Korobov2008}.



To address this discrepancy, here we consider the fully-coupled governing interaction equations that account for all triads satisfying PSI resonance condition \eqref{007}. We solve this governing equation through multiple-scale analysis that obtains a uniformly-valid nonlinear instability solution. Our analysis determines that, contrary to linear stability theory prediction, it is a pair of subharmonic waves with frequencies \textit{different from} $\omega_0/2$ that grow the largest in the PSI. In fact, in cases pairs of frequency $\omega_0/2$ receive the least amount of energy compared to other pairs in the pool of interactions. 

Furthermore, internal waves with frequencies near $\omega_0/2$, as can be derived from equation \eqref{007}, have very large vertical wavenumbers and therefore are very short. 
In fact, at exactly $\omega_0/2$, the vertical wavenumber is infinite. Strongest resonant waves of PSI, as predicted by the nonlinear stability theory, have frequencies far from $\omega_0$ hence have finite wavenumbers. Therefore, nonlinear stability theory shows that the actual strongest resonant waves of PSI have much larger scales than what linear stability theory predicts. 

While it is not unexpected that nonlinear theory results in a long-term growth which is different from the linear theory predictions, it is obvious that both linear and nonlinear theories must give the same initial growth rates. But, to our surprise, our initial growth rate from nonlinear stability theory did not match the prevalent linear growth rate reported in the literature. After some scrutiny we realized that  in the classical theory the effect of the second exponential term has been mistakenly neglected, resulting in an incorrect reported initial growth rate. Specifically, in linear stability analysis, amplitude growth in a resonance is usually obtained in the form $A=a\e^{bt}+c\e^{-bt}$, and therefore the growth rate is $(1/A)(\d A/\d t)$ = $ b(a\e^{bt}-c\e^{-bt})/(a\e^{bt}+c\e^{-bt})$. While at large times (at which linear stability analysis is usually not valid) the growth rate tends to $b$, the initial growth rate is $b[1-2c/(a+c)]$, that depends on $a,c$ and can potentially be very much different from $b$.



In what follows, we present our nonlinear stability analysis and discuss both initial growth rate and overall long-time growth of each resonance triads. We validate our analytical solution with direct numerical simulation results obtained from non-hydrostatic Navier-Stokes solver MITgcm.\\

\section{Interaction Equations for Parametric Subharmonic Instability}

Consider an inviscid, incompressible and stably-stratified fluid bounded by a top rigid lid and a flat seafloor at the bottom. In a Cartesian coordinates system with $x,y$-axes on the rigid lid and $z$-axis positive upward, the governing equations read
\bsa{901}
&\rho_0 \D\b{u}/\D t=-\g p-\rho g \g z,~~-h<z<0,\label{g101}\\
&\D\rho/\D t=0, ~~-h<z<0,\label{g1031}\\
&\g\cdot\b{u}=0, ~~-h<z<0,\label{g1032}\\
&w=0,~~z=0, \label{g1051}\\
&w=0,~~z=-h,\label{g1052}
\esa
where  $\b{u} = \{u, v, w\}$ is the velocity vector, $\rho$ is the density, $p$ is the pressure, and $g$ is the gravitational acceleration. A linear density profile is considered such that the background density is given by $\bar{\rho}(z)/\rho_0=1-a z$, with $\rho_0=\bar{\rho}(z)|_{z=0}$. Equation \eqref{g101} is the momentum equation, \eqref{g1031} represents conservation of salt (assuming that the density only depends on salinity in the equation of state and diffusion of salt is negligible), \eqref{g1032} is conservation of mass, and equations \eqref{g1051},\eqref{g1052} are kinematic boundary conditions on the rigid lid and the sea bottom respectively.

\begin{figure}
\centering
\includegraphics[width=9cm]{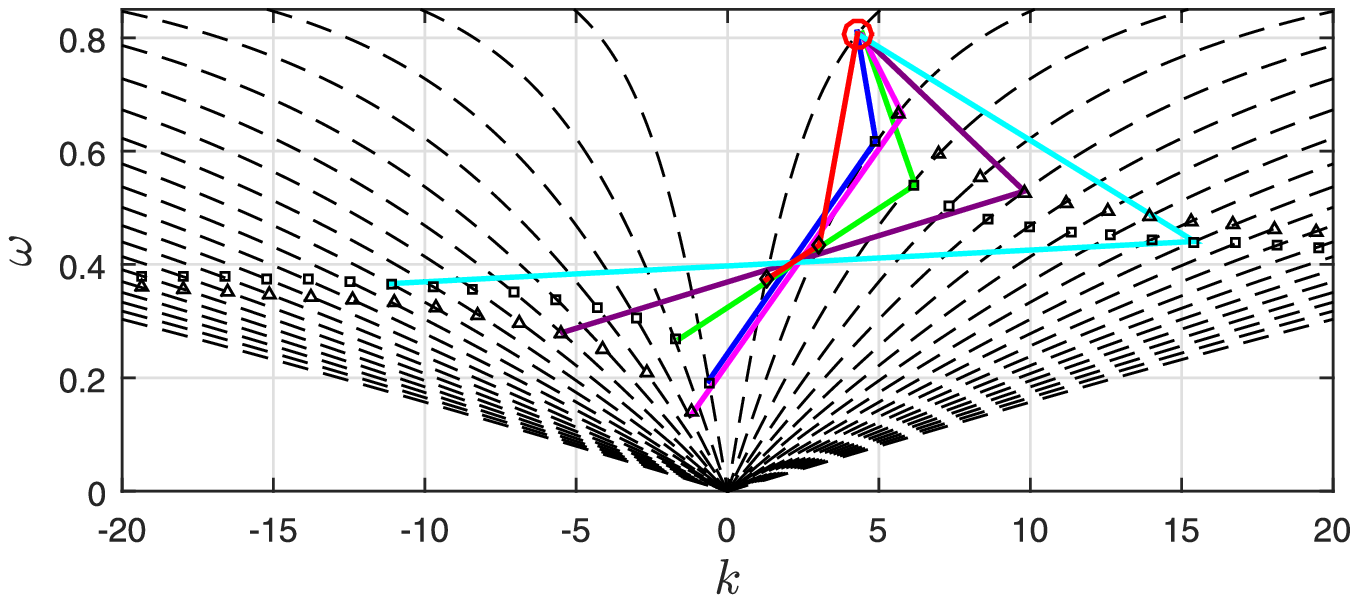}
\put(-245,82){(a)}\\
\includegraphics[width=9cm]{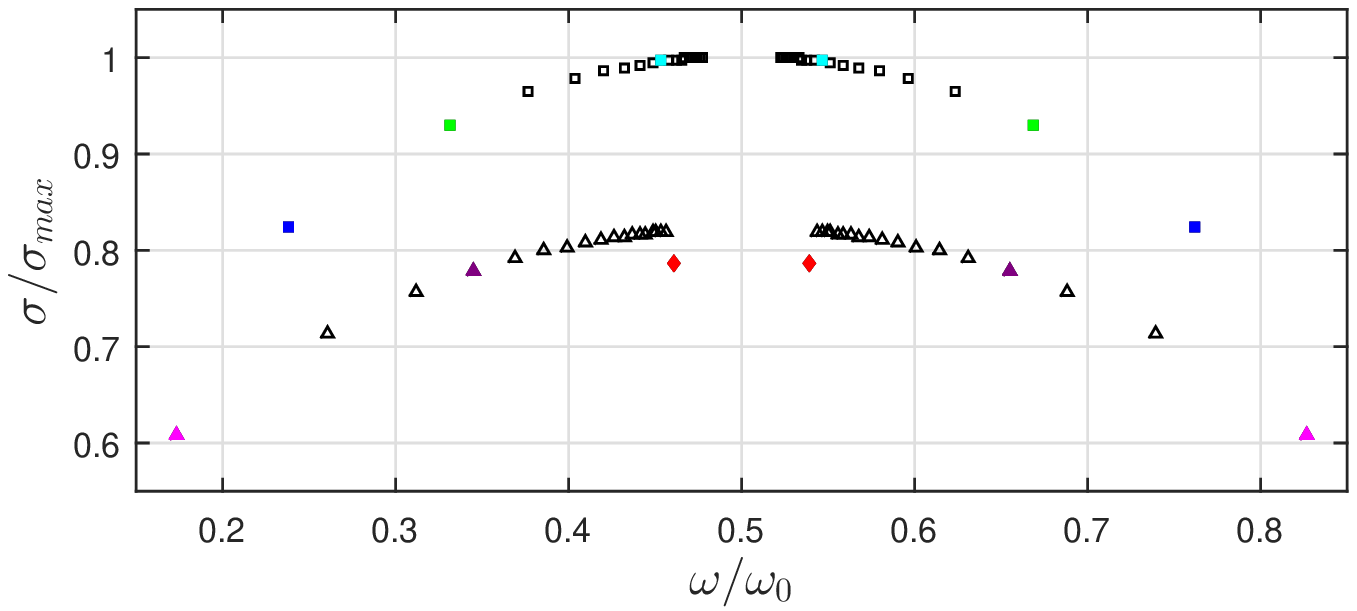}
\put(-245,82){(b)}\\
\includegraphics[width=9cm]{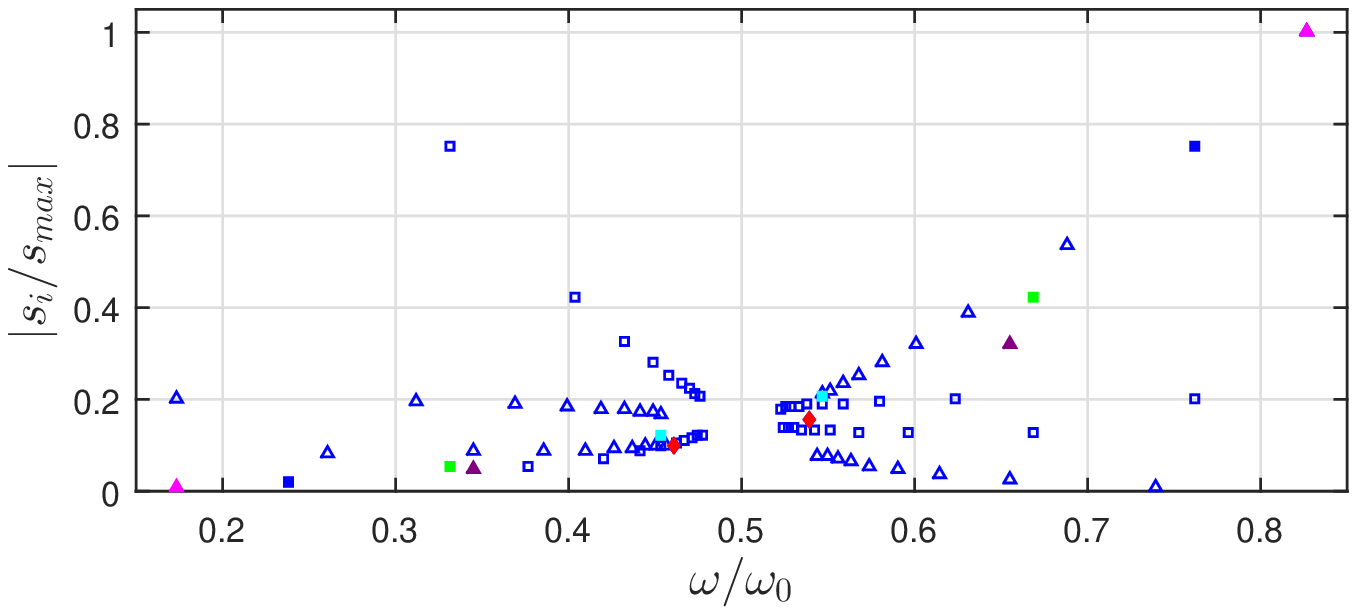}
\put(-245,82){(c)}\\
	\caption{(a) Geometric construction of several of possible triad resonances (c.f. \eqref{007}) between a primary internal wave of $\omega$=0.8 (red circle) and pairs of lower harmonic waves for $ah$=0.05. Square and triangles correspond to the higher mode propagating respectively in the same/opposite direction of the primary wave, and diamonds show a special case in which both subharmonic waves travel in the same direction. Six triads are highlighted by  color solid lines. Black dashed lines show branches of the dispersion relation $\omega=|k/\sqrt{k^2+m^2}|$. (b) Variation of the normalized exponent ($\sigma/\sigma_{max}$) of linear stability analysis \eqref{1452} as a function of normalized frequency $\omega/\omega_0$. Clearly subharmonic waves with frequencies close to $\omega/2$ have the highest exponents \cite[c.f. figure 11b of][]{Bourget2013}. (c) Normalized interaction coefficient $|s_i/s_{max}|$ as a function of normalized frequency $\omega/\omega_0$. The maximum initial growth rate occurs at the frequency $\omega/\omega_0$=0.83 although $\sigma$ is maximum at $\omega/\omega_0$=0.5.}
\label{fig11}
\end{figure}

System of equations \eqref{901} admits, among other solutions, a propagating internal-wave solution. Considering that a primary internal wave ($\k_0,\omega_0$) with a finite initial amplitude co-exists with two perturbation waves ($\k_1,\omega_1$) and ($\k_2,\omega_2$) the vertical velocity to the leading order can be written in the form  
\ba\label{131}
w=\sum_{j=0,1,2}A_j \sin m_j(z+h)e^{ i(k_j x-\omega_j t)}+\text{c.c.}
\ea
where $\k_j=k_j\hat i + m_j \hat z$, and \text{c.c.} denotes the complex conjugates. 
If triad resonance condition \eqref{007} is satisfied between the three waves, then amplitudes $A_j$ will slowly change with time. Mathematically this is expressed by $A_j=A_j(\epsilon t)$, that is, amplitudes are functions of \textit{slow time} ($\epsilon$ is a small parameter and a measure of the waves' steepness).


Let's first define the following dimensionless variables 
\ba
&\nn A^*_j=\f{A_j}{A_{00}}, ~~t^*=\f{t}{T_0}\cdot\f{A_{00}T_0}{h}, ~~\omega^*_j=\f{\omega_j}{N},~~ k^*_j=k_jh,~~m^*_j=m_jh.
\ea
where $A_{00}=A_0(t)|_{t=0}$. Through multiple-scale perturbation analysis, the differential equation that governs the evolution of a wave triad can be obtained, dropping all asterisks, as
\bsa{145}
\d A_0(t)/\d t=s_0 A_1(t)A_2(t)\\
\d A_1(t)/\d t=s_1 A_0(t)\bar{A}_2(t)\\
\d A_2(t)/\d t=s_2 A_0(t)\bar{A}_1(t)
\esa
where $\bar{A}_{1,2}(t)$  are complex conjugates  of $A_{1,2}(t)$, and\footnote{Our equation \eqref{146} is in fact the same as (3.26) presented in \cite{Bourget2013} except for the pre-factor $1/4$ which is due to \cite{Bourget2013} considering a vertically unbounded domain whereas in our study we have vertical wall boundaries. Therefore, in our case, unlike \cite{Bourget2013}, we must have standing waves in the vertical direction and also can only have discrete vertical wavenumbers $m$.}
\ba\label{146}
s=\frac{(k_1m_2-k_2m_1)}{4k_0k_1k_2} \lp \f{k_1}{\omega_1}-\f{k_2}{\omega_2} \rp \lp \f{k_0}{\m_0} +  \f{k_1}{\m_1}+\f{k_2}{\m_2}\rp \omega_0^2.
\ea
%
%
%
The interaction coefficients for the two perturbing waves $s_1$ and $s_2$ can be obtained by simply swapping the physical parameters in the expression for $s_0$ in \eqref{146} \cite[c.f. e.g.][]{Mcewan2006,Bourget2013}.

\section{Dominant Subharmonic Waves}

The subharmonic waves generated in the process of PSI must satisfy the resonant condition \eqref{007} and the dispersion relation $\omega=|k/\sqrt{k^2+m^2}|$. 
As an example of the wave triads satisfying the resonance conditions, for the choice of $ah=5\times10^{-2}$ and $\omega_0=0.8$, we present in figure \ref{fig11}a several of possible subharmonic wave triads that can be excited by PSI of the primary wave (red circle). The total number of triad possibilities is (countably) infinite. Note that there are two distinct branches of triads in figure \ref{fig11}a, marked by triangles and squares. Clearly, as the wavenumber of perturbation waves involved in the triad increases the frequency of perturbation waves asymptotically tend to half the frequency of the primary wave ($\omega_0/2$). In other words, perturbation waves with frequencies near $\omega_0/2$ have large horizontal and vertical wavenumbers, i.e., $|k_1,m_1|\approx|k_2,m_2|\gg|k_0,m_0|$ \cite[c.f. e.g.][]{Staquet2002}.


Based on the linearized instability theory, i.e. assuming that the amplitude of the primary wave is constant (of course this only applies for the very initial period of the resonance), then \eqref{145} becomes,
\bsa{1451}
d A_1(t)/dt=s_1 A_0\bar{A}_2(t),\\
d A_2(t)/dt=s_2 A_0 \bar{A}_1(t),
\esa
where $A_0$ is a constant. If the initial amplitudes of the two perturbing waves are respectively $A_1|_{t=0}=\delta_1$ and $A_2|_{t=0}=\delta_2$, the solution to $A_1(t)$ is,
\ba\label{1452}
A_1(t)=&1/2(\f{s_1  A_0 \bar{\delta_2}}{\sigma}+\delta_1) \exp(\sigma t)\nn\\
&-1/2(\f{s_1 A_0 \bar{\delta_2}}{\sigma}-\delta_1) \exp(-\sigma t)
\ea
where $\sigma=\sqrt{s_1s_2} |A_0|$. A similar expression is obtained for $A_2(t)$ with subscripts 1,2 in \eqref{1452} swapped.

In classical theory of PSI, $\sigma$ has been considered as the measure of growth of perturbation waves. 
We show in figure \ref{fig11}b the plot of $\sigma/\sigma_{max}$ (i.e. $\sigma$ normalized by the maximum $\sigma$ found from all possible PSI triads) as a function of $\omega/\omega_0$, where $\omega$ refers to the frequency of perturbation waves. Colors of markers in figure \ref{fig11}b are associated with the colors of triads depicted geometrically in figure \ref{fig11}a. 
In each resonance, two perturbation waves are involved that are shown with a pair of same-color markers. 
For instance, triad of green color occurs between perturbation waves of frequencies $\omega_{1,2}/\omega_0$= 0.34, 0.66, and results in $\sigma/\sigma_{max}$=0.93. 
The two arc-shaped branches formed in this plot correspond to the two branches of triad possibilities in figure \ref{fig11}a (triangles and squares, as discussed before). 
Behavior of $\sigma$ presented in figure \ref{fig11}b matches exactly figure 11b of \citep{Bourget2013} except that in our case because of two horizontal top and bottom boundaries we get discrete modes only. 

\begin{figure}
\centering
\includegraphics[width=9cm]{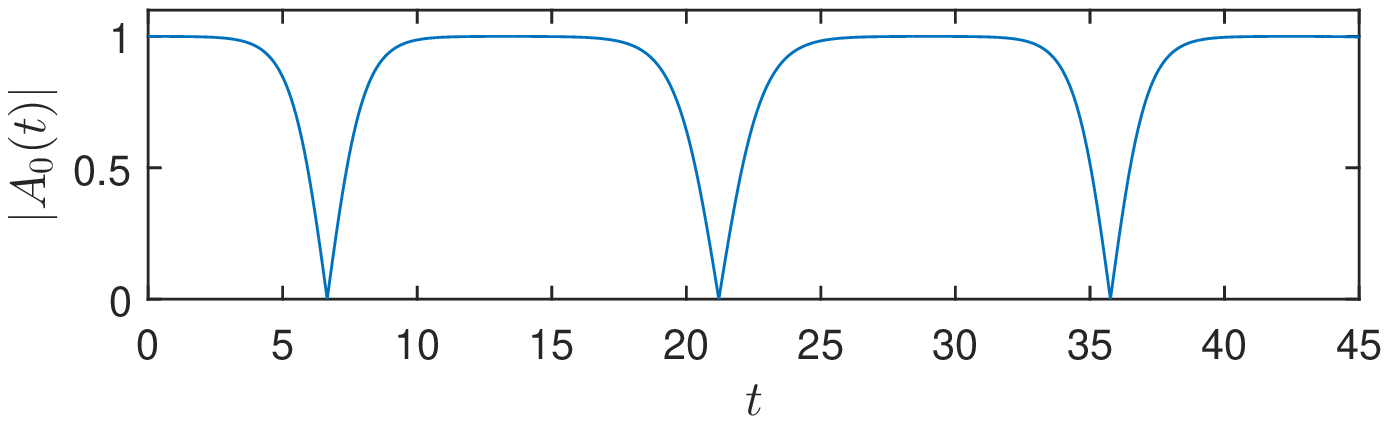}
\put(-260,70){(a)}\\
\includegraphics[width=8cm]{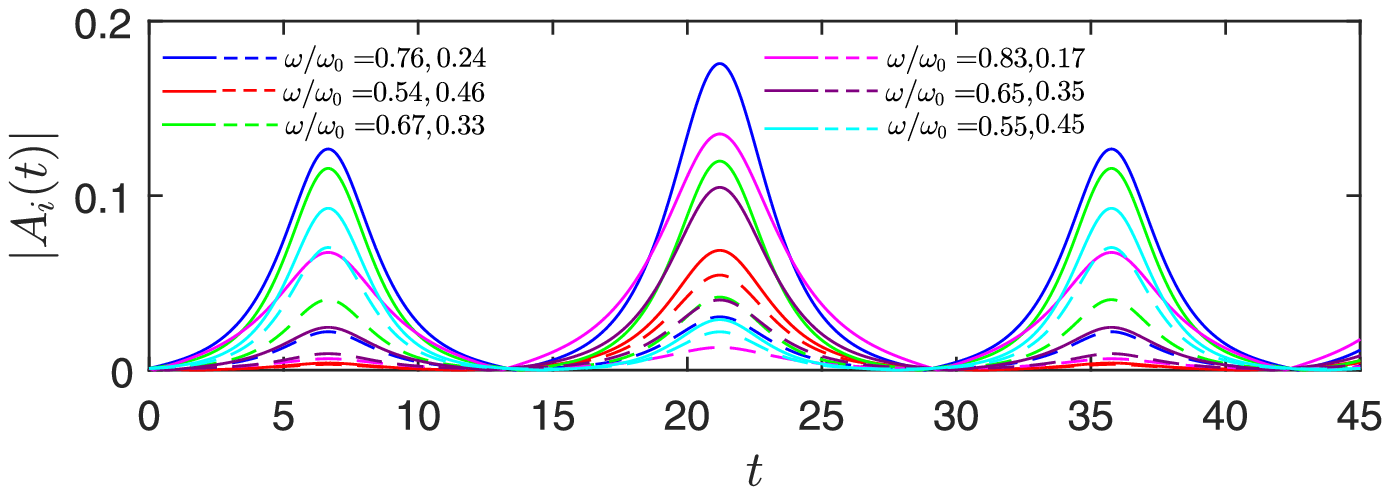}
\put(-240,75){(b)}
\caption{Evolution of amplitudes of (a) the primary internal wave and (b) the six subharmonic pairs undergoing PSI corresponding to the case represented in figure \ref{fig11}. Initial amplitudes of perturbation waves chosen as $A_{i1/i2}|_{t=0}$=0.001 and up to mode 20th (which results in 38 PSI pairs) are considered in the numerical integration of \eqref{155}. All waves undergo a modulation in time, and the maximum growth is obtained for perturbation wave of $\omega/\omega_0$=0.76. 
}
\label{fig21}
\end{figure}


Clearly figure \ref{fig11}b suggests that the maximum of $\sigma$ is at $\omega/\omega_0$=0.5. But it is to be noted that this does not imply that the initial growth rate nor the long-time growth is highest at $\omega/\omega_0$=0.5. Specifically, at large $t$, it is in fact expected that the exponential term dominates the behavior. However, the linear stability analysis is not valid for large $t$, and a nonlinear analysis must be called. The linear theory is only applicable at initial times, and the initial growth rate is given by
\ba\label{500}
\f{dA_{1,2}(t)}{\d t} \bigg\rvert_{t=0}=s_{1,2}A_0\bar\delta_{2,1}.
\ea
Effect of $\sigma$ in the exponent, as can be seen from \eqref{1452}, is canceled by the coefficients in front, and therefore, the initial growth rate \eqref{500}, is determined by $s_{1,2}$, and not $\sigma$. Behavior of normalized correct initial growth rate $s_i/s_{max}$ plotted against $\omega/\omega_0$ (figure \ref{fig11}c) shows almost an opposite behavior when compared with that of $\sigma$. Most importantly, the highest value of correct initial growth rate $s_{max}$, does not occur at $\omega/\omega_0$=0.5, but at  $\omega/\omega_0$=0.85. In fact, perturbation waves with frequencies near $\omega/\omega_0$=0.5 have almost the smallest initial growth rates\footnote{We would like to note that $s_i$ for some of the perturbation waves is negative and therefore the amplitude will first decrease. Usually in such cases, since the amplitude of perturbation waves are initially small, the amplitude quickly decreases to zero and then starts to grow on the negative side. The negative amplitude means that the wave gains a $\pi$-radian phase difference.}.

To determine the long term growth of waves involved in PSI, nonlinear stability analysis must be conducted. Since all pairs of perturbation waves (that satisfy resonance condition) simultaneously interact with the primary wave, a correct from of \eqref{145} that takes into account the coupling between waves read 
\bsa{155}
&\d A_0(t)/\d t=\sum_{i=1}^{N}s_{i0} A_{i1}(t)A_{i2}(t)\\
&\d A_{i1}(t)/\d t=s_{i1} A_0(t)\bar{A}_{i2}(t)\\
&\d A_{i2}(t)/\d t=s_{i2} A_0(t)\bar{A}_{i1}(t)
\esa
where subscript $i$ denotes the $i$th subharmonic wave pair. From \eqref{155}, it can be rigorously shown that 
\ba\label{1551}
\f{\d}{\d t} \bigg\{ \f{|A_0|^2}{\omega_0^2}+\sum_{i=1}^{N}\bigg[\f{|A_{i1}|^2}{\omega_{i1}^2} +\f{|A_{i2}|^2}{\omega_{i2}^2}\bigg] \bigg\} =0,
\ea
which shows that our governing equation \eqref{155} conserves energy. In other words, although energy is exchanged between a large number of waves, the total energy of the system is conserved and does not change with the time.

As a case study, consider a primary internal wave of $\omega_0=0.8$ in a stratified water of $ah=5\times10^{-2}$ which is correspond to figure \ref{fig11}. If we consider wave modes up to the mode 20th, that is 20 branches of the dispersion relation plot (dashed lines) in figure \ref{fig11}a, then 38 pair of perturbation waves can be found to form resonance with our primary wave (6 of them are shown in figure \ref{fig11}a). We assume all these perturbation waves have the same initial amplitude $A_i$=$1\times 10^{-3}$, which corresponds to a white-noise-like distribution of perturbation waves in the environment. Results of long time evolution of the primary wave, as well as few important perturbation waves, are shown in figure \ref{fig21}a,b: at the stage $t<$6.8, amplitudes of some of perturbation waves increase, in cases substantially and even by orders of magnitudes, at the expense of a decrease in the energy of primary wave. Once the entire energy of the primary wave is depleted, then the process reverses and now energy flows back from (initially) perturbation waves to the primary wave. This modulation continues with a period of $T_i\sim$ 15 for primary wave and with the period of $T_p\sim 2T_i$ for perturbation waves.  

The most important feature of long-time evolution plots (figure \ref{fig21}b) is the fact that the largest growth is observed at frequency $\omega/\omega_0=$0.76 ($A_i=$0.18 at t=21.7). The second and third largest growth are respectively at frequency $\omega/\omega_0=$0.83 ($A_i=$0.13 at t=21.7) and $\omega/\omega_0=$0.67 ($A_i=$0.12 at t=21.7); none at the frequency $\omega/\omega_0=$0.5. The perturbation wave associated with $\omega/\omega_0=$0.55 (which is closest to 0.5 in our database) gains a maximum value of $A_i=0.093$ which is barely 50\% of the highest growth.

\begin{figure}
\centering
\includegraphics[width=9cm]{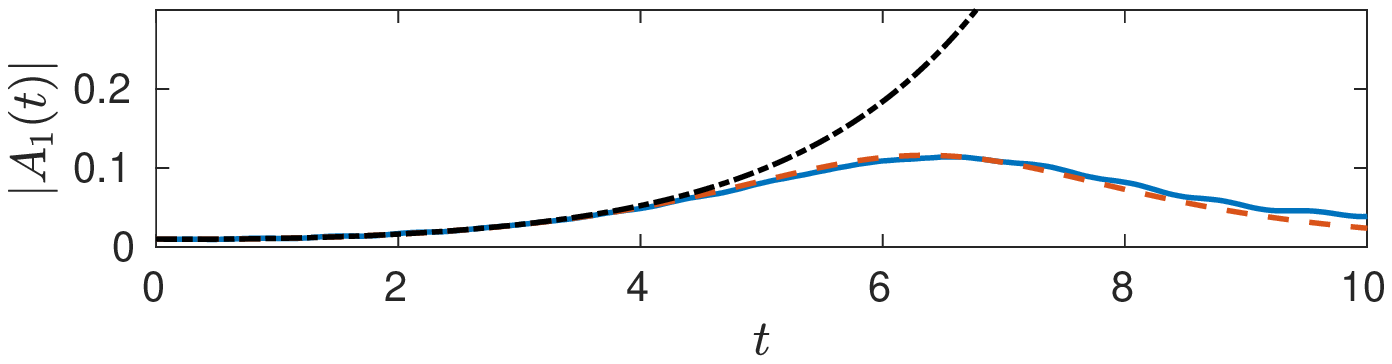}
\includegraphics[width=9cm]{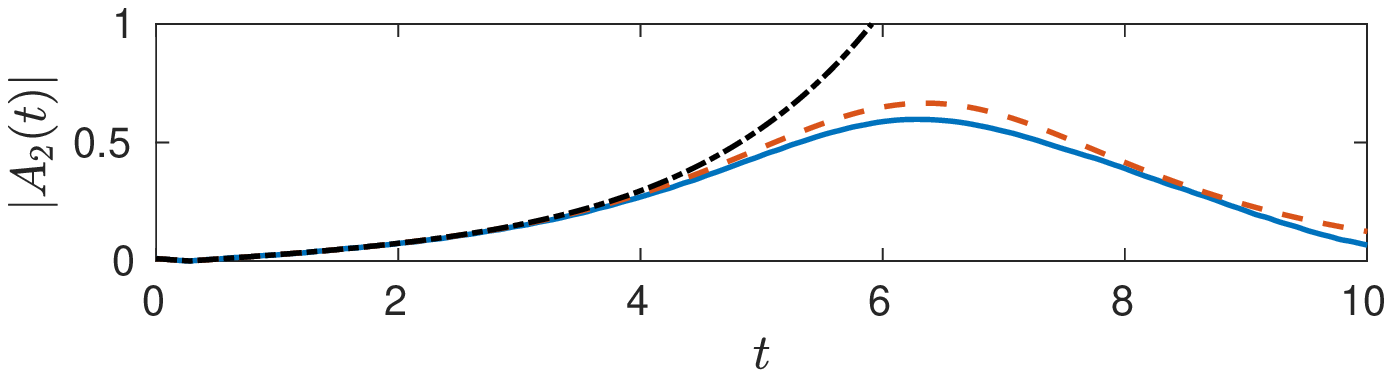}
\includegraphics[width=9cm]{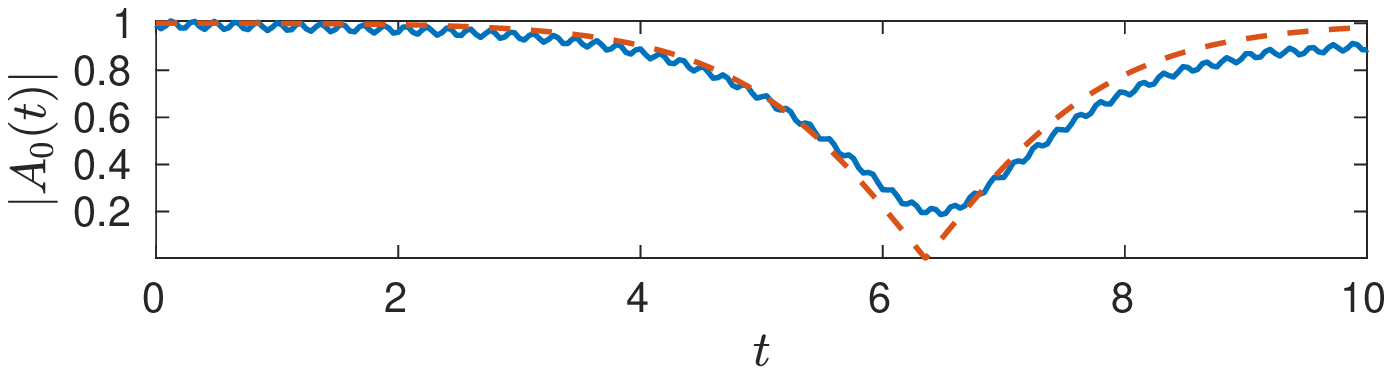}
	\caption{Time-evolution of a primary internal wave of frequency $\omega=0.8$ with two perturbation waves of frequencies $\omega_{1,2}/\omega_0$=0.24,0.76. Direct simulations results of MITgcm (blue solid line) matches well with our analytical results using nonlinear stability theory though numerical integration of \eqref{155} (red dashed line). For comparison, growth rate prediction by linear stability theory (black dash-dotted line) are also shown. Physical parameters for MITgcm simulations are $a=5\times10^{-5}$m$^{-1}$, $h=1000$m and initial amplitude is $1$m for the primary wave and $0.01$m for each of the two subharmonic waves.}
\label{fig6}
\end{figure}

We validate our analytical predictions with direct simulation run by MITgcm \cite[][]{Marshall1997}. MITgcm is a finite-volume based open-source non-hydrostatic Navier-Stokes solver that is widely used for modeling stratified mediums \cite[e.g.][]{Engqvist2004,Klymak2010,Lim2010,Churaev2015,Alford2015b}. We consider a rectangular domain with a periodic boundary condition on both ends in the x-direction, a free-slip wall at the bottom, and a free-slip rigid lid at the top. We use physical parameters $a=5\times10^{-5}$m$^{-1}$, $h=1000$m and consider a primary wave of mode one with $\omega_0=0.806$, $A_0|_{t=0}$=1m, that resonates two subharmonic waves of frequencies $\omega_{1,2}/\omega_0=$0.24, 0.76,  $A_{1,2}|_{t=0}$=$0.01$m\footnote{This specific pair of subharmonics are intentionally chosen since their wavenumber are $k_{1,2}/k_0$=-1/7,8/7 allowing us to use a periodic domain in the x-direction.}.
The evolution of amplitude of the three waves over time obtained from the direct simulation of MITgcm (blue solid lines in figure \ref{fig6}) compares well with our analytical results obtained from \eqref{155} (red dashed curves). For comparison, we also show the results of linear stability analysis (black dash-dotted lines) on top of the other two curves. Clearly, at the initial stage of resonance and as long as the change in the amplitude of $A_0$ is small, linear stability analysis estimates the growth of perturbations with a satisfactory tolerance. But for later stages, linear theory over-predicts the growth.

\section{Conclusion}

In summary, we showed, contrary to widely-accepted results drawn from linearized instability theory, that Parametric Subharmonic Instability (PSI) is strongest among a primary internal waves and perturbation waves at frequencies different from half the frequency of internal wave. Specifically, we proved that both the initial growth rate and the maximum amplitude reached by the (initially) perturbation waves are highest for subharmonic waves of low vertical modes with frequencies higher than $\omega_0/2$. It is straightforward to show that similar conclusion holds also in the spatial (or boundary value) problem, i.e. as waves propagate away from a source (e.g. a wavemaker) and interaction develops over the distance \cite[c.f. e.g.][]{alam2009bragg}. Our finding suggests that the efficiency of converting internal wave energy from large scales to small scales through PSI may have been overestimated by previous studies, and dominant resonant waves may have been missed if sought at the half frequency. 

\newpage
\bibliographystyle{unsrt}

\begin{thebibliography}{10}

\bibitem{wunsch2004}
Carl Wunsch and Raffaele Ferrari.
\newblock Vertical mixing, energy, and the general circulation of the oceans.
\newblock {\em Annu. Rev. Fluid Mech.}, 36:281--314, 2004.

\bibitem{Garrett2003a}
Chris Garrett.
\newblock {Internal tides and ocean mixing}.
\newblock {\em Science (New York, N.Y.)}, 301(5641):1858--9, 2003.

\bibitem{Boyd2007}
Philip~W. Boyd.
\newblock {Biogeochemistry: Iron findings}.
\newblock {\em Nature}, 446(7139):989--991, 2007.

\bibitem{Harris1986}
Graham~P. Harris.
\newblock Chapman and Hall, 1986.

\bibitem{Lamb2014}
Kevin~G Lamb.
\newblock {Internal Wave Breaking and Dissipation Mechanisms on the Continental
  Slope/Shelf}.
\newblock {\em Annual Review of Fluid Mechanics}, 46:231--54, 2014.

\bibitem{Sarkar2017}
S.~Sarkar and A.~Scotti.
\newblock From topographic internal gravity waves to turbulence.
\newblock {\em Annual Review of Fluid Mechanics}, 49(1):195--220, 2017.

\bibitem{Swinney2014}
Likun Zhang and Harry~L. Swinney.
\newblock Virtual seafloor reduces internal wave generation by tidal flow.
\newblock {\em Phys. Rev. Lett.}, 112:104502, 2014.

\bibitem{Hasselmann1967}
K.~Hasselmann.
\newblock {A criterion for nonlinear wave stability}.
\newblock {\em Journal of Fluid Mechanics}, 30(04):737, 1967.

\bibitem{Davis1967}
R~E Davis and A~Acrivos.
\newblock {The stability of oscillatory internal waves}.
\newblock {\em Journal of Fluid Mechanics}, 30(4):723--736, 1967.

\bibitem{Dauxois2013}
H\'el\`ene Scolan, Eugeny Ermanyuk, and Thierry Dauxois.
\newblock Nonlinear fate of internal wave attractors.
\newblock {\em Phys. Rev. Lett.}, 110:234501, 2013.

\bibitem{Lombard1996}
Peter~N Lombard and James~J Riley.
\newblock {Instability and breakdown of internal gravity waves. I. Linear
  stability analysis}.
\newblock {\em Physics of Fluids}, 8:3271--3287, 1996.

\bibitem{Karimi2014}
Hussain~H Karimi and T~R Akylas.
\newblock {Parametric subharmonic instability of internal waves: locally
  confined beams versus monochromatic wavetrains}.
\newblock {\em Journal of Fluid Mechanics}, 757:381--402, 2014.

\bibitem{Koudella2006}
C.~R. Koudella and C.~Staquet.
\newblock {Instability mechanisms of a two-dimensional progressive internal
  gravity wave}.
\newblock {\em Journal of Fluid Mechanics}, 548:165, 2006.

\bibitem{Dauxois2018}
Thierry Dauxois, Sylvain Joubaud, Philippe Odier, and Antoine Venaille.
\newblock {Instabilities of Internal Gravity Wave Beams}.
\newblock {\em Annual Review of Fluid Mechanics}, 50:1--28, 2018.

\bibitem{Phillip1981}
M~Phillip.
\newblock {Wave interactions - the evolution of an idea}.
\newblock {\em Journal of Fluid Mechanics}, 106:215--227, 1981.

\bibitem{Marcus2009}
Chung-Hsiang Jiang and Philip~S. Marcus.
\newblock Selection rules for the nonlinear interaction of internal gravity
  waves.
\newblock {\em Phys. Rev. Lett.}, 102:124502, 2009.

\bibitem{Alford2008b}
M.~H. Alford.
\newblock {Observations of parametric subharmonic instability of the diurnal
  internal tide in the South China Sea}.
\newblock {\em Geophysical Research Letters}, 35(15):2--6, 2008.

\bibitem{Chinn2012a}
Brian~S. Chinn, James~B. Girton, and Matthew~H. Alford.
\newblock {Observations of internal waves and parametric subharmonic
  instability in the Philippines archipelago}.
\newblock {\em Journal of Geophysical Research: Oceans}, 117(5):1--12, 2012.

\bibitem{Martin1972}
S.~Martin, W.~Simmons, and C.~Wunsch.
\newblock {The excitation of resonant triads by single internal waves}.
\newblock {\em Journal of Fluid Mechanics}, 53:17--44, 1972.

\bibitem{McEwan1971}
A.~D. McEwan.
\newblock {Degeneration of resonantly-excited standing internal gravity waves}.
\newblock {\em Journal of Fluid Mechanics}, 50(03):431--448, 1971.

\bibitem{Bourget2013}
Baptiste Bourget, Thierry Dauxois, Sylvain Joubaud, and Philippe Odier.
\newblock {Experimental study of parametric subharmonic instability for
  internal plane waves}.
\newblock {\em Journal of Fluid Mechanics}, 723:1--20, 2013.

\bibitem{Staquet2002}
C.~Staquet and J.~Sommeria.
\newblock {Internal gravity waves : from instabilities to turbulence}.
\newblock {\em Annual Review of Fluid Mechanics}, 34:559--593, 2002.

\bibitem{Joubaud2012}
Sylvain Joubaud, James Munroe, Philippe Odier, and Thierry Dauxois.
\newblock {Experimental parametric subharmonic instability in stratified
  fluids}.
\newblock {\em Physics of Fluids}, 24(24):41703--23116, 2012.

\bibitem{Korobov2008}
Alexander~S. Korobov and Kevin~G. Lamb.
\newblock {Interharmonics in internal gravity waves generated by
  tide-topography interaction}.
\newblock {\em J . Fluid Mech}, 611:61--95, 2008.

\bibitem{Mcewan2006}
A.~D. Mcewan, D.~W. Mander, and R.~K. Smith.
\newblock {Forced resonant second-order interaction between damped internal
  waves}.
\newblock {\em Journal of Fluid Mechanics}, 55(04):589, 1972.

\bibitem{Marshall1997}
J.~Marshall, A.~Adcroft, C.~Hill, L.~Perelman, and C.~Heisey.
\newblock {A finite-volume, incompressible Navier Stokes model for studies of
  the ocean on parallel computers}.
\newblock {\em Journal of Geophysical Research}, 102:5753--5766, 1997.

\bibitem{Engqvist2004}
A.~Engqvist and A.M. Hogg.
\newblock {Unidirectional stratified flow through a non-rectangular channel}.
\newblock {\em J. Fluid Mech.}, 509:83--92, 2004.

\bibitem{Klymak2010}
J.~M. Klymak, S.~M. Legg, and R.~Pinkel.
\newblock {High-mode stationary waves in stratified flow over large obstacles}.
\newblock {\em Journal of Fluid Mechanics}, 644:321--336, 2010.

\bibitem{Lim2010}
K.~Lim, G.~N. Ivey, and N.~L. Jones.
\newblock {Experiments on the generation of internal waves over continental
  shelf topography}.
\newblock {\em Journal of Fluid Mechanics}, 663:385--400, 2010.

\bibitem{Churaev2015}
E.~N. Churaev, S.~V. Semin, and Y.~A. Stepanyants.
\newblock {Transformation of internal waves passing over a bottom step}.
\newblock {\em Journal of Fluid Mechanics}, 768:1--11, 2015.

\bibitem{Alford2015b}
Matthew~H. Alford, Thomas Peacock, and Jennifer~A. MacKinnon~et al.
\newblock {The formation and fate of internal waves in the South China Sea}.
\newblock {\em Nature}, 521(7550):65--69, 2015.

\bibitem{alam2009bragg}
Mohammad-Reza Alam, Yuming Liu, and Dick~KP Yue.
\newblock Bragg resonance of waves in a two-layer fluid propagating over bottom
  ripples. part ii. numerical simulation.
\newblock {\em Journal of Fluid Mechanics}, 624:225--253, 2009.

\end{thebibliography}

\end{document}